\documentclass[pre,twocolumn,showpacs,aps,floats,floatfix,superscriptaddress]{revtex4}

\usepackage{graphicx}
\usepackage{subfigure}
\usepackage{amsmath, amsthm, amssymb}

\begin{document}

\newcommand{\avk}{\langle k \rangle}
\newcommand{\fluck}{\langle k^2 \rangle}

\title{Co-evolution of density and topology in a simple model
  of city formation}

\author{Marc Barth\'elemy} \affiliation{CEA-Centre d'Etudes de
  Bruy{\`e}res-le-Ch{\^a}tel, D\'epartement de Physique Th\'eorique et
  Appliqu\'ee BP12, 91680 Bruy\`eres-Le-Ch\^atel, France}
\affiliation{Centre d'Analyse et Math\'ematique Sociales (CAMS, UMR
  8557 CNRS-EHESS), Ecole des Hautes Etudes en Sciences Sociales, 54
  bd. Raspail, F-75270 Paris Cedex 06, France}
\author{Alessandro Flammini}
\affiliation{School of Informatics, 
Indiana University, 901 E. Tenth st., 47408, Bloomington, IN}

\date{\today} 
\widetext

\begin{abstract}
  
  We study the influence that population density and the road network
  have on each others' growth and evolution.  We use a simple model of
  formation and evolution of city roads which reproduces the most
  important empirical features of street networks in cities. Within
  this framework, we explicitely introduce the topology of the road
  network and analyze how it evolves and interact with the evolution
  of population density. We show that accessibility issues -pushing
  individuals to get closer to high centrality nodes- lead to high
  density regions and the appearance of densely populated centers. In
  particular, this model reproduces the empirical fact that the
  density profile decreases exponentially from a core district. In
  this simplified model, the size of the core district depends on the
  relative importance of transportation and rent costs.

\end{abstract}

%\pacs{89.75.-k, -87.23.Ge, 05.40.-a}

%89.75.-k       Complex systems
%87.23.Ge       Dynamics of social systems
%05.40.-a Fluctuation phenomena, random processes, noise, and Brownian motion

\maketitle 

%%%%%%%%%%%%%%%%%%%%%%%%%%%%%%% intro: modeling cities
\section{Introduction}

%%%%%   importance of the subject

It has been recently estimated that more than $50\%$ of the world
population lives in cities and this figure is bound to increase
\cite{UN}. The migration towards urban areas has dictated a fast and
short-term planned urban growth which needs to be understood and
modelled in terms of socio-geographical contingencies, and of the
general forces that drive the development of cities. Previous studies
\cite{Christaller,Levinson,Fujita} about urban morphology have mostly
focused on various geographical, historical, and social-economical
mechanisms that have shaped distinct urban areas in different ways.  A
recent example of these studies can be found in~\cite{Levinson}, where
the authors study the process of self-organization of transportation
networks with a model that takes into account revenues, costs and
investments.  

The goal of the present study is to model the coupling between the
evolution of the transportation network and the population
density. More precisely, the question we aim to answer is the
following. Given the pattern of growth of the entire population of a
given city, how is the local density of population changing within the
boundaries of the city itself, and how the road network's topology is
modified in order to accommodate these changes?  There are in
principle a huge number of potentially relevant factors that may
influence the growth and shape of urban settlements, first and
foremost the social, economical and geographical conditions that
causes the population of a given city to increase in a particular
moment of its history. We neglect in the present study this class of
factors and consider the overall growth in the number of inhabitants
as an exogenous variable.  In order to achieve conclusions that have a
good degree of generality, and, at the same time, to maintain the
number of assumptions as limited as possible, we focus on two main
features only: the local density of population and the structure of
the road network. Population density and the topology of the network
constitute two different facets of the spatial organization of a city,
and from a purely qualitative point of view it is not hard to believe
that their evolution is strongly correlated.  Indeed, Levinson, in a
recent case study~\cite{Levinson2} about the city of London in the
$19^{th}$ and $20^{th}$ centuries has demonstrated how the changes in
population density and transportation networks deployment are strictly
and positively correlated.  Obviously, the road network tends to
evolve to better serve the changing density of population. In turn,
the road network influences the accessibility and governs the
attractiveness of different zones and thus, their growth. However,
attractiveness leads to an increase in the demand for these zones,
which in turn will lead to an increase of prices. High prices will
eventually limit the growth of the most desirable areas. It is the
mutual interaction between these processes that we aim to model in the
present work.

Although there are many other economical mechanisms (type of land use,
income variations, etc.) which govern the individual choice of a
location for a new `activity' (home, business, etc), we limit
ourselves to the two antagonist mechanisms of accessibility and
housing price. These loosely defined notions can be taken into account
when translated in term of transportation and rent costs. We note that
in the context of the structure of land use surrounding cities, von
Th\"unen~\cite{vonthunen:1966} already identified the distance to the
center (a simple measure of accessibility) and rent prices as being
the two main relevant factors.

At first we will discuss separately the two mechanisms of road formation and
location choice. In particular, we explicitely consider the
shape of the network and model its evolution as the result of a local
cost-optimization principle~\cite{Barthelemy:2007}.  In classical
models used in urban economics, transportation costs are usually
described in a very simplified fashion in order to avoid the
description of a separate transportation industry~\cite{Fujita}. Also,
when space is explicitly taken into consideration, the shape of the
transportation networks is rarely considered and transportation costs
are computed according to the distance to a city center (as it is the
case in the classical von Th\"unen's \cite{vonthunen:1966} or
Dixit-Stiglitz's \cite{Dixit:1977} models). In these
approaches transportation networks are absent, and displacements of
goods and individuals are assumed to take place in continuous
space. On one side this allows for a more detailed description of the
economical processes at play during the shaping of a city. On the
other side, these approaches often rely on the hypothesis that the processes shaping
a city are slow enough to allow the balancing of the different forces
that contribute to these processes, allowing as a consequence the achievement 
of the global minimum of some opportune cost function. 

The point of view inspiring our work, instead, is that the evolution
of a city is inherently an `out-of-equilibrium' process where the
city evolves in time to adapt to continuously changing
circumstances. If some sort of optimization or `planning' is driving the
growth, it has to be continuously redefined in order to take into account the
ever-changing economic and social conditions that are ultimately
responsible for the evolution of urban areas. We do not, therefore,
assume the optimization of a global cost (or utility) function.

Finally, we would like to mention that our goal is not to be as
realistic as possible but to consistently reproduce a set of coarse
grained and very general features of real cities under a minimal set
of plausible assumptions.  Alternative explanations might also be
possible and it would be interesting to compare our results with those
produced in the same spirit. We hope that this simplified model could serve as a
first step in the direction of designing more elaborated models.
 
%%%%% organization of the paper

This paper is organized in three main parts. In the first part, we
briefly establish the framework to describe the model and discuss the
empirical evidences that motivated it. In the second part, we
address the issue of how the growth of the local density
affects the growth of the road network. In the third part we will
study how the road network affects the potential for density growth in
different areas. We finally integrate all these elements in the fourth
section, where we study the full model and discuss our results.

\section{The model: empirical evidences and definition}
% --------- empirical measures and discussion
\subsection{Framework}

In our simplified approach, we represent cities as a collection of
points scattered on a two dimensional area (a square of linear size
$L$ throughout this study), and connected by a urban road network. The
description of the street network adopted here consists of a graph
whose links represent roads, and vertices represent roads'
intersections and end points. Although the primary interest here is on
roads' networks~\cite{Cardillo1,Buhl}, 

it is worth mentioning that transportations networks appear in variety
of different contexts including plant/leaves
morphology~\cite{Rolland}, rivers~\cite{Iturbe}, mammalian circulatory
systems~\cite{West}, commodity delivery~\cite{Gastner}, and
technological infrastructures~\cite{Schwartz}. Indeed, networks are
the most natural and possibly simplest representation of a
transportation system ~\cite{stevens,ball}.  It would be impossible to
review, even schematically, the approaches and the insights gained in
the specific fields mentioned above, but it is at least worth to
mention a few studies which attempted to connect the evolution of
networks to an optimization principle (as it is the case of the
present work).  Maybe is it not surprising that man-made
transportation networks have been designed with the goal to serve
efficiently and cost little~\cite{Gastner}, but relevant examples
occur in natural sciences as well. It is remarkable, for example, how
the Kirchoff law, that determines the current in the edges of a
resistor network can be derived assuming the minimization of the
dissipated energy~\cite{doyle}. More recently, optimization principles
have been successfully applied to the study of the transportation of
nutrients through mammalian circulatory systems in order to explain
the allometric scaling laws in biology~\cite{West,Banavar}.  River
networks constitute a further example where relevant features of the
network organization can be derived from an optimality
principle~\cite{Iturbe,maritan}. A last example worth mentioning is
that of metabolic networks, where it has been found that specific
pathways appear if conditions for optimal growth are assumed (see {\it
  e.~g.}~\cite{price}). It is interesting to notice that there have
been attempts to put some of the examples discussed above in the
context of a single framework (see~\cite{fittest}).  In addition, let
us mention that there is a huge mathematical literature that studies
optimal networks and the flow they support; Minimal Spanning
Trees~\cite{mst}, Steiner Trees~\cite{steiner}, and Minimum Cost
Network Flows~\cite{mcfn} are just three examples.  Although the
present study assumes a notion of optimality, there are some important
aspects that differentiate it from the works discussed above. The
first is that the principle of optimality is at work only locally:
there is no global cost function that our road networks are supposed
to minimize. The second, and possibly more important, is that we
attempt to establish a connection between the evolution of the network
and that of the quantity that such network is supposed to transport,
i.e. population.  Transportation networks, as shown from the example
cited above, can generally display a large variety of
patterns. However, recent empirical studies
~\cite{Batty,Makse1,Makse2,Crucitti,Jiang,Cardillo1,Cardillo2,Lammer,Porta,Roswall:2005,Jiang:2004}
have shown that roads' networks, despite the peculiar geographical,
historical, social-economical processes that have shaped distinct
urban areas in different ways, exhibit unexpected quantitative
similarities, suggesting the possibility to model these systems
through quite general and simple mechanisms. In the following
subsection we present the evidences that support the previous
statement.

\subsection{Empirical results}

The degree distribution of planar networks decays very fast for large
degrees as it usually happens for networks embedded in Euclidean space
or with strong physical constraints which prohibits the emergence of
hubs~\cite{Amaral:2000}. The degree distribution is therefore strongly
peaked around its average (over the whole city) $\langle k \rangle=2E/N\equiv 2e$ ($E$ is
the number of edges-the roads-and $N$ is the number of nodes-the
intersections). Concerning the average degree of random planar
networks, little is known: For one- and two-dimensional lattices $e=1$
and $e=2$, respectively, and a classical result shows that for any
planar network $\langle k\rangle\leq 6$, implying $e\leq 3$ (see
{\it e.g.} \cite{Itzykson}). It has also been recently shown that
planar networks obtained from random Erdos-Renyi graphs over a
randomly plane-distributed set of points upon rejection of non-planar
occurrences, have $e>13/7$~\cite{Gerke}.
\begin{figure}[t!]
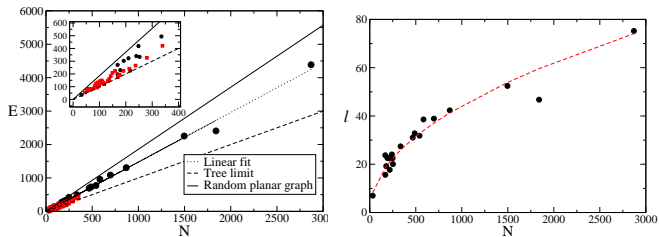

   \centering
   \includegraphics[angle=0,scale=.17]{fig1a.eps}
   \includegraphics[angle=0,scale=.17]{fig1b.eps}
   \caption{Left: Numbers of roads versus number of nodes
     (ie. intersections and centers) for data from
     \protect\cite{Cardillo1} (circles) and from \cite{Buhl}
     (squares). In the inset, we show a zoom for a small number of
     nodes. Right: Total length versus the number of nodes. The line
     is a fit which predicts a growth as $\sqrt{N}$ (data from
     \cite{Cardillo1}).}
   \label{fig:k_cost}
\end{figure}
These facts are summarized in Fig.~\ref{fig:k_cost} (left), together
with empirical data from $20$ cities in different
continents~\cite{Cardillo1}. A first important empirical observation is
that $1.05\leq e_{emp}\leq 1.69$ , in a range lying between trees and
2d lattices, and the average degree over all these cities is $\langle
k\rangle\approx 2.87$.  The strongly peaked degree distribution
suggests that a quasi-regular lattice could give a fair account of the
road network topology.  This suggestion is reinforced if one considers
the cumulative length $\ell$ of the roads. With this picture in mind,
one would expect that for a given average density $\rho=N/L^2$, the
typical inter distance between nodes is $\ell_1\sim
\frac{1}{\sqrt{\rho}}$. The total length is then the number of edges
times the typical inter-distance which leads to
\begin{equation}
\ell\sim E\ell_1\sim \frac{\langle k\rangle}{2}L\sqrt{N}
\end{equation}
This behavior is reproduced in fig.~(\ref{fig:k_cost}b), where a fit
of the empirical data in \cite{Cardillo1} with a function of the form
$aN^{1/2}$ gives $a_{emp}\approx 1.46\pm 0.04$ (a fit  with a function of the form
$aN^{\tau}$ leads to $a\approx 1.51\pm 0.24$ and $\tau\approx 0.49\pm 0.03$). The
value $a_{emp}=1.46$ has to be compared with the average degree over all
cities. One finds $2.87/2\approx 1.44$ and considering statistical
errors, it is hard to reject the hypothesis of a slightly perturbed
lattice as a model for the road network.

The two empirical facts above lend credibility to the simple picture
that city streets are described by a quasi-regular lattice with an
essentially constant degree (equal to approximately $3$) and constant
road length ($\ell_1\sim 1/\sqrt{\rho}$). There is however a further
empirical fact which forces us to reconsider this simple picture. The
roads' network define a tessellations of the surface and the authors
of \cite{Lammer} measured the distribution of the area $A$ of the
polygons delimited by the edges of the network. Surprisingly, they found
a power law behavior of the form
\begin{equation}
P(A)\sim A^{-\alpha}
\end{equation}
with $\alpha\simeq 1.9$ (the standard error is not available in
\cite{Lammer}). This fact contradicts the simple model of an almost
regular lattice since the latter would predict a distribution $P(A)$
very peaked around a value of the order of $\ell_1^2$.  The authors of
~\cite{Lammer} also measured the distribution of the form factor given
by the area of a cell divided by the area of the circumscribed circle
(for this value they use the largest distance $D$ between nodes of the
cell, a convention that we adopted): $\phi=4A/(\pi D^2)$. They found
that most cells have a form factor between $0.3$ and $0.6$ indicating
a large variety of cell shapes.

A first challenge is therefore to design a model for planar networks
that can reproduce quantitatively these featurees and which is based
on a plausible (small) set of assumptions. The simple indicators
discussed above show that one cannot model the network by either
lattices, Voronoi tessellation, random planar Erdos-Renyi graphs, all
these networks having a peaked distribution of areas and form
factors. Let's note that the scale-invariant distribution for cell
sizes can be obviously reproduced by assuming by the fractal model of
~\cite{Kalapala:2006} which assumes a self-similar process of road
generation. The power law distribution for cell sizes automatically
follows from this assumption. In the following, we present a model
that relies on a simple plausible mechanism, does not assume
self-similarity and quantitatively accounts for the empirical facts
presented above.

% --------- The venation model

\section{The model of road formation}

We first discuss the part of our model that describes the evolution of
the road network. Our main assumption is that the network grows by
trying to connect to a set of points -the `centers'- in an efficient
and economic way. These centers can represent either homes, offices or
businesses. This parameter free model is based on a principle of local
optimality and has been proposed in \cite{Barthelemy:2007}. For the
sake of self-consistency and readability, we first describe this model
in detail. The application of optimality principles to both natural
and artificial transportation networks has a long
tradition~\cite{Stevens,Ball}. The rationale to invoke a local
optimality principle in this context is that every new road is built
to connect a new location to the existing road network in the most
efficient way~\cite{Bejan}. During the evolution of the street
network, the rule is implemented locally in time and space. This means
that at each time step the road network is grown by looking only at
the current existing neighboring sites. This reflects the fact that
evolution histories greatly exceed the time-horizon of planners. The
self-organized pattern of streets emerges as a consequence of the
interplay of the geometrical disorder and the local rules of
optimality. In this regard our model is quite different from
approaches to transportation networks where an equilibrium situation
is assumed and which are based on either (i) minimization of an
average quantity ({\it e.g.} the total travel time), or (ii) on the
inclusion of many different socio-economical factors ( {\it e.g.}
land use).

\subsection{Network growth} 

\begin{figure}[!t]
  \vspace*{0.5cm}
   \centering
   \includegraphics[angle=0,scale=.30]{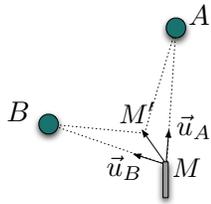}
   \caption{The nearest road to the centers $A$ and $B$ is $M$. The
     road will grow to point M'. The proposed minimum expenditure
     principle suggests that the next point M' will be such that the
     variation of the total distance to the two points A and B is
     maximal.}
   \label{fig:delta}
\end{figure}

When new centers (such as new homes or businesses) appear, they need
to connect to the existing road network.  If at a given stage of the
evolution a single new center is present, it is reasonable to assume
that it will connect to the nearest point of the existing road
network. When two or more new centers are present (as in
Fig.~\ref{fig:delta}) and they want to connect to the same point in
the network, we assume that economic considerations impose that a
single road - from the chosen network's point - is built to connect both
of them.
In the example of figure~\ref{fig:delta}, the nearest point of the
network to both new centers $A$ and $B$ is $M$. We grow a single new
portion of road of fixed length $dx$ from $M$ to a new point $M'$ in
order to grant the maximum reduction of the cumulative distance of $A$
and $B$ from the network. This translates in the requirement that
\begin{equation}
\delta d=d(M,A)+d(M,B)-[d(M',A)+d(M',B)]
\end{equation}
is maximal ($dx$ being fixed). A simple calculation shows that the
maximization of $\delta d$ leads to
\begin{equation}
d\overrightarrow{MM'}\propto \vec{u}_A+\vec{u}_B
\label{vec_rule}
\end{equation}
where $\vec{u}_A$ ($\vec{u}_B$) is the unitary vector from $M$ to $A$
($B$).

%\footnote{}. 

The procedure described above is iterated until the road from $M$
reaches the the line connecting $A$ and $B$, where a singularity
occurs: $d\overrightarrow{MM'}=0$.  From there two independent roads to $A$ and
$B$ need to be built to connect to the two new centers.  The rule
Eq.~(\ref{vec_rule}) can be easily generalized to the case of $n$ new
centers, and, interestingly, was proposed in the context of
visualization of leafs' venation patterns~\cite{Rolland}.

The growth scheme described so far leads to tree-like structures and
we implement ideas proposed in~\cite{Rolland} in order to create
networks with loops. Indeed, even if tree-like structures are
economical, they are hardly efficient: the length of the path along a
minimum spanning tree network for example, scales as a power $5/4$ of
the Euclidean distance between the end-points~\cite{Duco}. Better
accessibility is then granted if loops are present. In order to obtain
loops, we assume, following~\cite{Rolland}, that a center can affect
the growth of more than one single portion of road per time step and
can stimulate the growth from any point in the network which is in its
relative neighborhood, a notion which has been introduced
in~\cite{Toussaint}. In the present context a point $P$ in the network
is in the relative neighborhood of a center $C$ if the intersection of
the circles of radius $d(P,C)$ and centered in $P$ and $C$,
respectively, contains no other centers or point of the
network~\cite{Toussaint}. This definition rigorously captures the
loosely defined requirement that, for $v$ to belong to the relative
neighborhood of $s$, the region between $s$ and $v$ must be empty. At
a given time step, a generic center $C$ then stimulates the addition
of new portions of road (pointing to P) from all points in the network
that are in its relative neighborood, naturally creating loops. When
more than one center stimulates the same point P the prescription
of~(\ref{vec_rule}) is applied and the evolution ends when the list of
stimulated points is exhausted (We refer the interested reader to
\cite{Barthelemy:2007} for a detailed exposition of the algorithm).

The formula above can be straigthforwardly extended to the case of
centers with non-uniform weight $\eta$. This leads to a modified version of
Eq.~(\ref{vec_rule}), where the sum of distances to be minimized is
weighted by $\eta$ and leads to
\begin{equation}
d\overrightarrow{MM'}\propto \eta_A\vec{u}_A+\eta_B\vec{u}_B  .
\label{vec_rule2}
\end{equation}
where $\eta_A$ and $\eta_B$ can be different.  Simulations with
non-uniform centers weights show that - as far as the location of
`heavy' and `light' centers is uniformly distributed in space,
uncorrelated and not broad - that the structure of the network is
locally modified, but that its large scale properties are virtually
unchanged.  In the algorithm presented above, once a center is reached
by all the roads it stimulates, it becomes inactive.  An interesting
variant of this model assumes that centers can stay active
indefinetively. In this case we expect a larger effect of the weights'
heterogeneity. We will leave this problem for future studies.

In the following, we study networks resulting from the growth process
described above. We assume that the appearance of new centers is given
exogenously and is independent from the existing road network and from
the position and number of the centers already present. The model
accounts quantitatively for a list of descriptors -the ones discussed
above in an empirical context- that characterize at a coarse grained
level the topology of street patterns. At a more qualitative level,
the model leads to the presence of perpendicular intersections, and
also reproduces the tendency to have bended roads even if geographical
obstacles are absent.

\begin{figure}[t!]
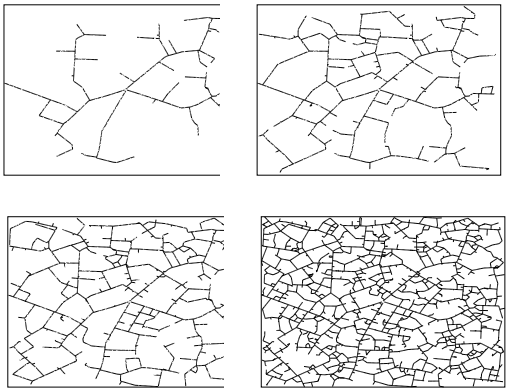

  \centering
  \begin{minipage}[b]{8 cm}
    \includegraphics[angle=0,scale=.15]{fig3a.eps}
    \includegraphics[angle=0,scale=.15]{fig3b.eps}
    \vspace{5 mm}
  \end{minipage}
  \begin{minipage}[b]{8 cm}
   \includegraphics[angle=0,scale=.15]{fig3c.eps}
   \includegraphics[angle=0,scale=.15]{fig3d.eps}
  \end{minipage}
  \caption{Snapshots of the network at different times of its
    evolution: for (a) $t=1,000$, (b) $t=2,000$, (c) $t=3,000$, (d)
    $t=4,000$. At short times, we have a tree structure and loops
    appear for larger density values obtained at larger times. For
    $t=4,000$, we have approximately $1,700$ nodes connected by $2,000$
    roads.}
   \label{fig:time}
\end{figure}

We show in Fig.~\ref{fig:time} examples of patterns obtained at
different times. The model gives information about the time
evolution of the road network: at earlier times, the density is low
and the typical inter-distance between centers is large (see
Fig.~\ref{fig:time}).  As time passes, the density increases and
the typical length to connect a center to the existing road network
becomes shorter. Since the number of points grows with time, the
simple assumption that the typical road length is given by
$1/\sqrt{\rho}$ leads to $\ell_1\sim 1/\sqrt{t}$ which is indeed what
the model predicts.

\begin{figure}[t!]
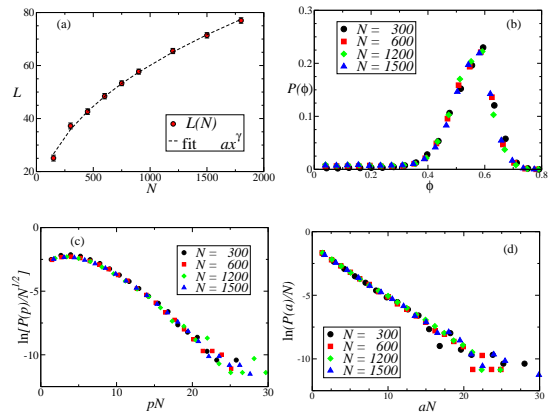

   \begin{center}
\subfigure{
   \includegraphics[angle=0,scale=.14]{fig4a.eps}
   \includegraphics[angle=0,scale=.14]{fig4b.eps}
}
\vspace{0.1cm}
\subfigure{
   \includegraphics[angle=0,scale=.14]{fig4c.eps}
   \includegraphics[angle=0,scale=.14]{fig4d.eps}
}
   \end{center}
   \caption{Results of the model (averaged over $1000$
     configurations). (a) Total length of roads versus the number of
     nodes. The dotted line is a square root fit. (b) Structure factor
     distribution showing a good agreement with the empirical results
     of~\protect\cite{Lammer}. (c-d) Rescaled distributions of the
     perimeter (c) and of the areas (d) of the cells displaying an
     exponential behavior.}
   \label{fig:l_phi}
\end{figure}

Beyond visual similarities, the model allows quantitative comparisons
with the empirical findings The ratio $e=E/N$, initially close to $1$
(indicating that the corresponding network is tree-like), increases
rapidly with $N$, to reach a value of order $1.25$ which is in the
ballpark of empirical findings. The cumulative length of the roads
produced by the model (Fig.~\ref{fig:l_phi}a) shows a behavior of the
form $a\sqrt{N}$ with $a\approx 1.90$, in good agreement with the
empirical measurements $a_{emp}\approx 1.87$). The form factor
distribution (Fig.~\ref{fig:l_phi}b) has an average value $\phi=0.6$
and values essentially contained in the interval $[0.3,0.7]$ in
agreement with the results in~\cite{Lammer} for 20 German cities.

\subsection{Effect of the center spatial distribution}
 
\begin{figure}[h!]
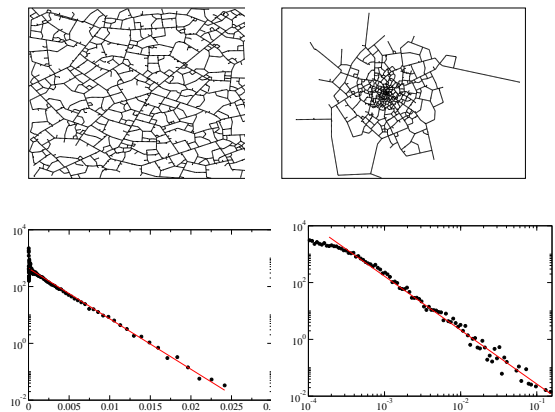

  \centering
  \begin{minipage}[b]{9.2 cm}
   \vspace{3 mm}
   \includegraphics[angle=0,scale=.15]{fig5a.eps}
   \includegraphics[angle=0,scale=.15]{fig5b.eps}
   \vspace{5 mm}
  \end{minipage}
  \begin{minipage}[b]{9.2 cm}
   \includegraphics[angle=0,scale=.15]{fig5c.eps}
   \includegraphics[angle=0,scale=.15]{fig5d.eps}
  \end{minipage}
  \caption{Upper left plot: Uniform distribution of points ($1000$
    centers, $100$ configurations). In this case, the area
    distribution is exponentially distributed (bottom left). Upper
    right plot: Exponential distribution of centers ($5000$ centers,
    $100$ configurations, exponential cut-off $r_c=0.1$). In this
    case, we observe a power law (bottom right). The line is a power
    law fit which gives an exponent $\approx 1.9$.}
   \label{fig:area}
\end{figure}

An important feature of street networks is the large diversity of cell
shapes and the broad distribution of cell areas. So far, we have
assumed that centers are distributed uniformly across the
plane. Within this assumption, the model predicts a cell area
distribution following an exponential (with a large cut-off however)
as shown in Fig.~\ref{fig:l_phi}(d) and Fig.~\ref{fig:area}.
The empirical distribution of centers in real cities, however, is not accurately
described by an uniform distribution but decreases exponentially from
the center~\cite{Makse1,Makse2}. We thus use such an exponential
distribution $P(r)=\exp (-|r|/r_c)$ for the center spatial location
and measure the areas formed by the resulting network (in the last
section of this point is further discussed). Although most
quantities (such as the average degree and the total road length) are
not very sensitive to the center distribution, the impact on the area
distribution is drastic. In Fig.~\ref{fig:area} a power law with
exponent equal to $1.9\pm 0.1$ is found, in remarkable agreement with the
empirical facts reported in \cite{Lammer} for the city of
Dresden. Although we cannot claim that this exponent is the same for
all cities, the appearance of a power law in good agreement with
empirical observations confirms the fact that the simple local
optimization principle is a possible candidate for the main process
driving the evolution of city street patterns. This result also
demonstrates that the centers' distribution is crucial in the
evolution process of a city.

The optimization process described above has several interesting
consequences on the global pattern of the street network when
geographical constraints are imposed, as illustrated by the following
example.  We simulated the presence of a river assuming that new
centers cannot appear on a stripe of given width (and are otherwise
uniformly distributed). 
\begin{figure}
   \centering
   \includegraphics[angle=0,scale=.25]{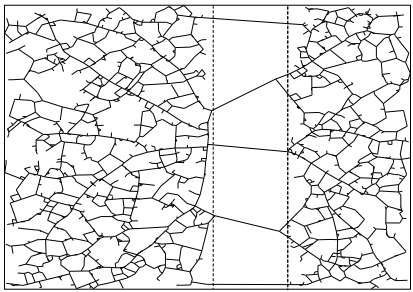}
   \caption{In the presence of an obstacle (here a `river'
     delimited by the two dotted line) in which the centers are not
     allowed to be located, the local optimization principle leads to
     a natural solution with a small number of bridges.}
   \label{fig:river}
\end{figure} 
The resulting pattern is shown in Fig.~\ref{fig:river}. The local
optimization principle naturally creates a small number of bridges
that are roughly equally spaced along the river and organizes the road
network. To conclude, it is worth noting that in the present
framework we didn't attempt the modelization of planning
efforts. Simulations show that, at the present simplified stage, the
presence of a skeleton of ``planned'' large roads has the effect of
partitioning the plane in different regions where the growth of the
network is dominated by the mechanism described above, and reproducing
on a smaller scale the structures shown in fig.~\ref{fig:time}.

\subsection{Hierarchical structure of the traffic} 

Finally, we discuss now the presence of hierarchy in the network
generated by the model. Indeed, geographers have recognized for a long
time (see {\it e.g.}~\cite{Christaller}) that many systems are
organized in a hierarchical fashion. Highways are connected to
intermediate roads which in turn dispatch the traffic through smaller
roads at smaller spatial scales. In order to test for the existence of
such a hierarchy in our model, we use the edge betweenness centrality
as a simple proxy for the traffic on the road network. For a generic
graph, the betweenness centrality $g(e)$ of an edge
$e$~\cite{Freeman,Goh:2001,Barthelemy:2003} is the fraction of
shortest paths between any pair of nodes in the network that go
through $e$. Allowing the possibility of multiple shortest paths
between two points, one has
\begin{equation}
g(e)=\sum_{s\neq t}\frac{\sigma_{st}(e)}{\sigma_{st}}
\end{equation}
where $\sigma_{st}$ is the number of shortest paths going from $s$ to
$t$ and $\sigma_{st}(e)$ is the number of shortest paths going from
$s$ to $t$ and passing through $e$. Central edges are therefore those
that are most frequently visited if shortest paths are chosen to move
from and to arbitrary points. 
\begin{figure}[t!]
   \begin{center}
\subfigure{
   \includegraphics[angle=0,scale=.30]{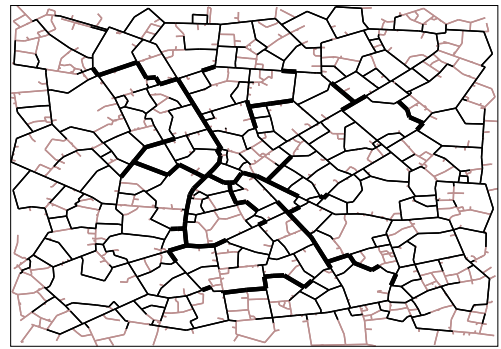}
}
   \end{center}
   \caption{Traffic map of the network. The edge centrality is
     computed and divided in three different groups and the thickness
     of the edge is plotted according to the group (from thin to thick
     for increasing edge centrality.}
   \label{fig:hierarchy}
\end{figure} 
We computed this quantity for all edges of the road network generated
by our algorithm. It appears that this quantity $g(e)$ is broadly
distributed and varies over more than $6$ orders of magnitude. In
order to get a simple representation of this quantity we arbitrarily
group the edges in three classes: $[1,10^4]$, $[10^4,10^5]$,
$[10^5,\infty]$ and plot them with different thicknesses. In
particular, we see in Fig~\ref{fig:hierarchy}, that edges with the
largest centrality (represented by the thickest line) form almost a
tree of large arteries. 
Proportionality between traffic and edge-centrality, as defined above,
is virtually equivalent to assuming: {\it i)} a uniform origin-destination
matrix, {\it ii)} everybody choose the shortest path to reach a destination, and
{\it iii)} roads are ``large'' enough to support the traffic generated by {\it i)} and 
{\it ii)} without congestion effects.   
Under these assumption one indeed observes a hierarchy of smaller roads
and streets with a decreasing typical length and the existence of a hierarchical 
structure of arteries, roads and streets.

\section{Location of centers: effect of density and accessibility}

In the simple version of the model presented above, the location of
centers are independent from the topology of the road network. In real
urban systems, this is however unlikely to happen. There is an
extensive spatial economics literature (see \cite{Fujita} and
references therein) that focuses on the several factors that may
potentially influence the choice location for new businesses, homes,
factories, or offices (see also \cite{Jensen} and references
therein). Empirical evidences suggest a strong correlation between
transportation networks and density increase have been recently
provided by Levinson~\cite{Levinson2}. Our goal here is to discuss,
based on very simple and plausible assumptions, the coupled evolution
of the road network and the population density.

We first divide the city in square sectors of area $S$, and we assume
that the choice of a location for a new center is governed by a
probability $P(i)$ that one ($i$) of these sectors is chosen. This
probability, which reflects the attractiveness of a location, depends
a priori on a large number of factors such as accessibility, renting
prices, income distribution, number and quality of schools, shops,
etc. A key observation made by previous authors (see for example
~\cite{Brueckner} and references therein) is that commuting cost
differences must be balanced by differences in living spaces
prices. We will follow this observation and we will thus focus on two
factors which are the rent price and the accessibility (which we will
reduce to commuting costs).

\subsection{Rent price and accessibility}

The housing price of a given location is probably determined by many
factors comprising tax policies, demography, etc., and is by itself an
important subject of study (see for example \cite{Goodman}). We will
make here the simplifying assumption that the rent price is an
increasing function of the local density (for each grid sector)
$\rho(i)=N(i)/S$, where $N(i)$ is the number of centers in the sector
$(i)$, and in particular that the rent price $C_R$ is directly
proportional to the local density of population (which can be seen as
the first term of an expansion of the price as a function of the
density)
\begin{equation}
C_R(i)=A\rho(i)
\end{equation}
where $A$ is some positive prefactor corresponding to the price per
density. We note here that a more general form of the type
$C_R=A\rho(i)^\tau$ could be used. A preliminary study suggests that as long 
as the rent cost is an increasing function of the density, our results remain 
qualitatively unchanged. It would be however very interesting to measure this function 
empirically.

The second important factor for the choice of a location is its
accessibility. Locations which are easily accessible and which allow
to reach easily arbitrary destinations are more attractive,
all other parameters being equal. Also, for a new commercial activity,
high traffic areas can strongly enhance profit opportunities. In terms
of the existing network, the best locations are therefore the most
central and standard models of city formation (see for example
\cite{Fujita}) indeed integrate the distance to the center and its
associated (commuting) cost as a main factor. Euclidean distance,
however, can be a poor estimator of the effective accessibility of a
given location, if this location is poorly connected to the
transportation network.  This is why the notion of centrality has to
be considered not only in geographical terms, but also from the point
of view of the network that grants mobility. The possibility to easily
reach an arbitrary location when movement is constrained by a network
is nicely captured in quantitative terms by the notion of node
betweenness centrality.

\begin{figure}[h!]
  \centering
  \begin{minipage}[b]{9.2 cm}
   \vspace{3 mm}
   \includegraphics[angle=0,scale=.35]{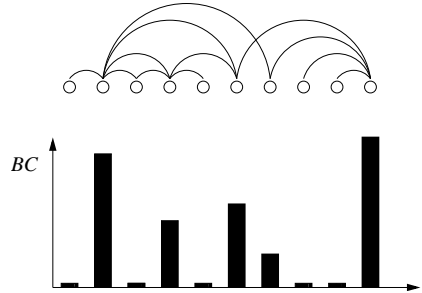}
  \end{minipage}
  \caption{Illustration of the notion of node betweenness centrality
    for a non planar graph}
   \label{fig:bw}
\end{figure}

In the previous section, we defined the {\it edge} betweenness
centrality and here we need a similar quantity defined for
nodes rather than for edges. The node betweenness centrality $g(v)$
\cite{Freeman,Goh:2001,Barthelemy:2003} of a node $v$ is defined
as the fraction of shortest paths between any pair of points in the
network that go through $v$. The mathematical expression of this
quantity is then
\begin{equation}
g(v)=\frac{1}{N(N-1)}\sum_{s\neq t}\frac{\sigma_{st}(v)}{\sigma_{st}}
\end{equation}
where $\sigma_{st}$ is the number of shortest paths going from $s$ to
$t$ and $\sigma_{st}(v)$ is the number of shortest paths going from
$s$ to $t$ and passing through $v$.  Betweenness centrality was
initially introduced as a natural substitute for geometric centrality
in graphs that are not embedded in Euclidean space (see fig.~\ref{fig:bw}).

\begin{figure}[h!]
  \centering
  \begin{minipage}[b]{9.2 cm}
   \vspace{3 mm}
   \includegraphics[angle=0,scale=.40]{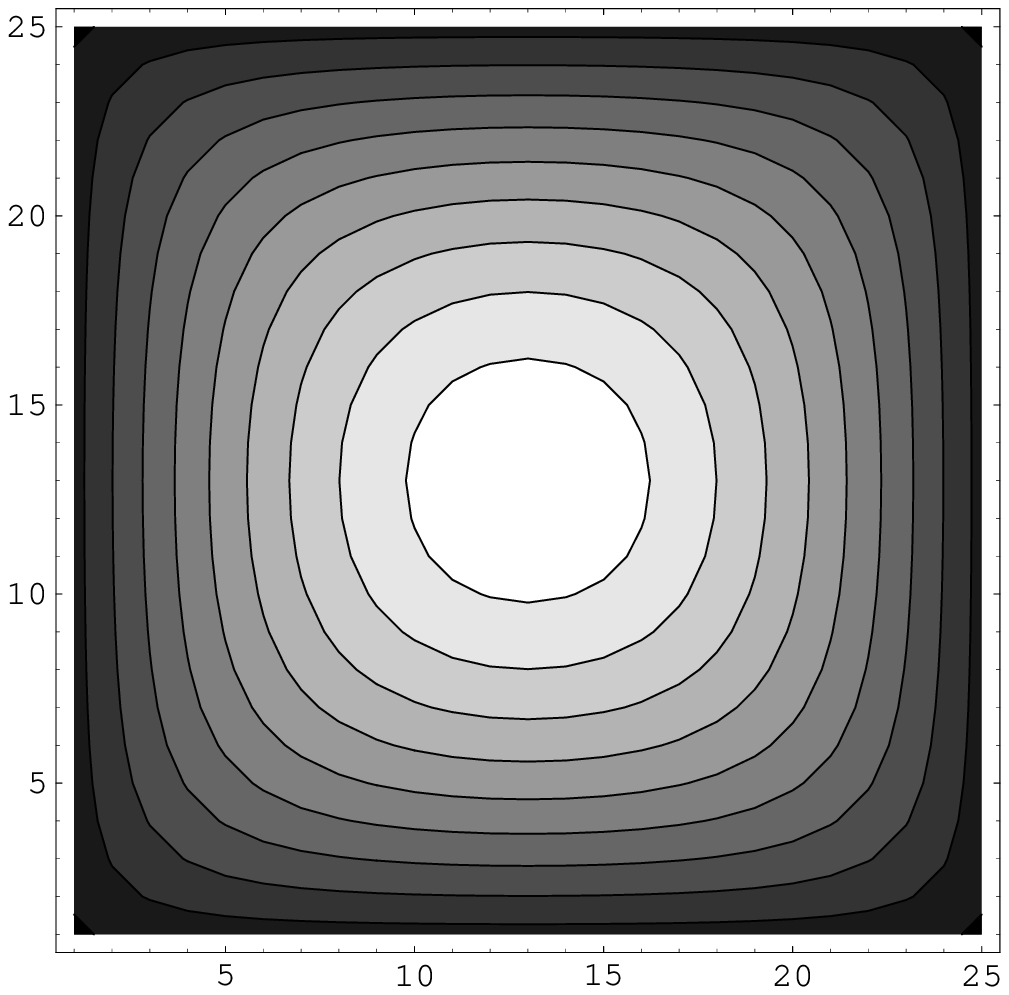}
   \includegraphics[angle=0,scale=.50]{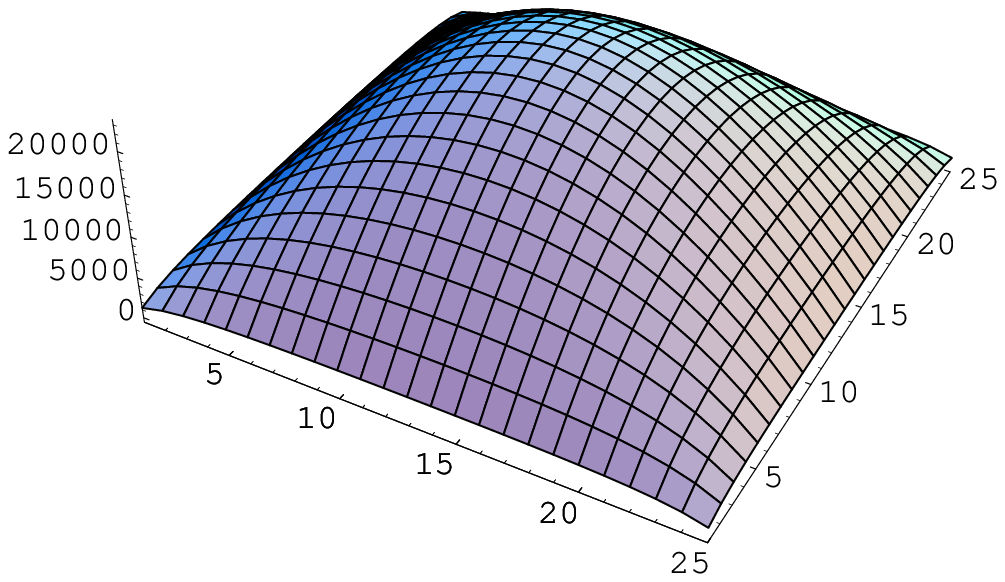}
   \vspace{5 mm}
  \end{minipage}
  \begin{minipage}[b]{9.2 cm}
   \includegraphics[angle=0,scale=.20]{fig9c.eps}
  \end{minipage}
  \caption{Top panel: Contour plot of node betweenness centrality -not
    normalized-for a square grid of linear size 25 (A light color
    indicates a high value of the betweenness).  Middle panel: 3D plot of node
    betweenness centrality (not normalized) for the same square
    grid. Bottom panel: Plot of the node betweenness of a generic node
    vs its average distance from all the other nodes.}
   \label{fig:bw2}
\end{figure}

Betweenness also naturally serves our purpose to quantify
accessibility on planar graphs, especially in our simplified framework
where an explicit distinction between resources and users has been
sacrified to the sake of simplicity.  It is important to note that
betweenness centrality, on planar graphs, is strictly correlated to
other, more common, measures of centrality.  The two first panels of
fig~\ref{fig:bw2} show the contour and the 3D plot of node-betweeness
for a Manhattan-like grid of 25 blocks per side and clearly show how
central nodes are those that are the most frequently visited if
shortest paths are chosen to move from and to arbitrary points.  The
third panel of the same figure shows the relation between the
betweenness of a node and its average distance from all other nodes
(along the same square grid). This plot demonstrates that the larger
the betweenness of a node is, the shorter is its average distance from
a generic node.
The betweenness thus extends the concept of geographical centrality to
networks whose structure is not a lattice or planar. 

In our simple model, we will assume that accessibility reduces here to
the facility of reaching quickly any other location in the
network. This can also be seen as the average commuting cost which in
previous models \cite{Fujita} was assumed to be proportional to the distance
to the center. The natural extension for a network is then to take the
transportation cost depending on the betweenness centrality. For each
sector $S_i$ of the grid, we first compute the average betweenness
centrality as
\begin{equation}
\overline{g}(i)=\frac{1}{N(i)}\sum_{v\in S_i}g(v)  .
\end{equation}
where the bar represents the average over all nodes (centers and intersections)
which belong in a given sector.

Transportation costs are a decreasing function of the
betweenness centrality and we will assume here that the transportation
cost $C_T(i)$ for a center in sector $(i)$ is given by
\begin{equation}
C_T(i)=B(g_m-\overline{g}(i))
\end{equation}
where $B$ and $g_m$ are positive constants (other choices, as long as
the cost decreases with centrality, linearly or not, give similar
qualitative results).

Finally, we will assume (as it is frequently done in many models, see
for example \cite{Brueckner} and references therein) that all new
centers have the same income $Y(c)=Y$. This assumption is certainly a
rough approximation, as demonstrated by effects such as urban
segregation, but in order to not overburden our model, we will neglect
income disparities in the present study. The net income of a new center
$c$ in a sector $(i)$ is then
\begin{equation}
K(i)=Y-C_R(i)-C_T(i).
\end{equation}
The higher the net income $K(i)$ and the more likely the location
$(i)$ will be chosen for the implantation of a new business, home,
etc. In urban economics the location is usually chosen by minimizing
costs, and we relax this assumption by defining the probability that a
new center will choose the sector $(i)$ as its new location under the form
\begin{equation}
P(i)=\frac{e^{\beta K(i)}}{\sum_je^{\beta K(j)}}  .
\end{equation}
This expression rewritten as
\begin{equation}
P(i)=\frac{  e^{\beta ( \lambda \overline{g}(i)-\rho(i)) } }
{\sum_j e^{\beta ( \lambda \overline{g}(j)-\rho(j)) } }
\label{eq:proba}
\end{equation}
where $\beta A$ is redefined as $\beta$ and where $\lambda=B/A$. For
numerical simulations, the local density is normalized by the global
density $N/L^2$, in order to have the density and centrality
contributions defined in the same interval $[0,1]$. The relative
weight between centrality and density is then described by $\lambda$
and the parameter $\beta$ implicitly describes in an `effective' way
all the factors (which could include anything from individual taste to
the presence of schools, malls, etc) that have not been explicitly taken into
account, and that may potentially influence the choice of location. If
$\beta\approx 0$, cost is irrelevant and new centers will appear
uniformly distributed across the different sectors:
\begin{equation}
P(i)\sim \frac{1}{N(i)}   .
\end{equation}
In the opposite case, $\beta\rightarrow\infty$, the location with the minimal
cost will be chosen deterministically.
\begin{equation}
\begin{cases} P(i)=1 & \text{for $i$ such that $K(i)$ is minimum}
\\
P(i)=0 & \text{for all other sectors}
\end{cases}
\end{equation}

The parameter $\beta$ can thus be used in order to adjust the
importance of the cost relative to that of other factors not explicitly
included in the model.

%%%%%%%%%%%%%%%%%%%%%%%%%%%%%%%%%%%%%%%%%%%%%%
\section{Co-evolution of the network and the density}

We finally have all the ingredients needed to simulate the
simultaneous evolution of the population density and the road network.
Before introducing the full model, analogously to what we have done
for the first part of the model, it is worth to study this second part
separately.  To do that, we consider a toy -one dimensional- case,
where the network plays no role, since a single path only exists between
each pair of nodes. Despite the simplicity of the setting, it is
possible to draw some general conclusion.

\subsection{One-dimensional model}

\begin{figure}
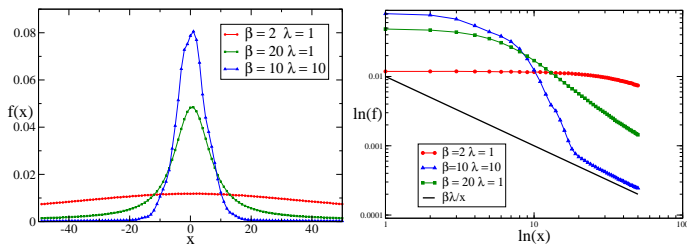

%\vspace*{.1cm} 
\centerline{
\includegraphics*[width=0.25\textwidth]{fig10a.eps}
\includegraphics*[width=0.25\textwidth]{fig10b.eps}
}
\caption{The stationary growth rate for different values of the
  parameters.(a) Large values of $\beta$ and $\lambda$ implies larger
  degree of centralization and a faster decay of density from center
  to periphery. (b) At large values of $\lambda$ the decay of density
  becomes algebraic for location away from the center. The exponent
  approaches $-1$ and $f(x)$ is approximated in that region by
  $1/(\beta \lambda /x)$.}
\label{oned}
\end{figure}

We assume that the centers are located on a one-dimensional segment
$[-L,L]$. Since only a single path exists between any two points, the
calculation of centrality is trivial. In the continuous limit, and for
a generic location $x$ it can be written as the product of the number
of points that lie at the right and left of the given location
\begin{equation}
g(x)=\int_{-L}^{x}\rho(y,t)dy \left[N-\int_{-L}^{x}\rho(y,t)dy\right]
\end{equation}
where $\rho(x,t)$ is the density at $x$. The equation for the
density therefore reads:
\begin{equation}
\partial_{t} \rho(x,t) =e^{\beta \left[ \lambda \frac{\int_{-L}^{x}\rho(y,t)dy}{N} 
\frac{( N-\int_{-L}^{x}\rho(y,t)dy)}{N}-\frac{\rho(x,t)}{N}\right]}
\label{eq:oned}
\end{equation}
where $N=\int_{-L}^{L}\rho(y,t)dy$.  The numerical integration
of Eq.~(\ref{eq:oned}) shows that, after a transient regime, the process locks in a
pattern of growth in which the total population grows at a constant
rate
\begin{equation}
N=\int_{-L}^{L}\rho(y,t)dy \propto t.
\end{equation} 
This suggests that a solution for large $t$ can be found via the
separation of variables under the form
\begin{equation}
\rho(x,t)=\alpha f(x) t
\label{eq:sepvar}
\end{equation}
where one can set $\int_{-L}^{L}f(x)dx=1$ without loss of generality.
Plugging the expression (\ref{eq:sepvar}) into Eq.~(\ref{eq:oned}) one gets

\begin{equation}
\alpha f(x)= 
e^{ 
\beta
\left[ 
\lambda \int_{-L}^{x}f(y)dy (1-\int_{-L}^{x}f(y)dy)-f(x)
\right]
}
\label{onedeq}
\end{equation}
where $\alpha$ is an integration constant to be determined.  An
explicit solution for the inverse $f^{-1}(x)$ can be achieved via the
Lambert function (Lambert's function is the principal branch of the
inverse of $z=w \exp^w$), but the expression is not particularly
illuminating and it is therefore not presented here.  Several facts
can however be understood using a direct numerical 
integration of Eq. (\ref{eq:oned}) or the simulation of the relative
stochastic process:
\begin{itemize}
\item{At large times, population in different location grows with a
    rate $f(x)$ that depends on the location but not explicitely on
    time.  This is a direct consequence of Eq.~(\ref{eq:sepvar}) and
    is obviously a different behavior from uniform growth. The ratio
    of population density in two different locations $x_1$ and $x_2$
    stabilizes in the long run to $f(x_1)/f(x_2)$}
  \item{Although $\beta$ models the `noise' in the choice of location
      and $\lambda$ the relative importance of centrality as compared
      to density, they have similar effects on the expected density in
      a given location. An increase in $\beta$ and $\lambda$
      corresponds a concentration of density in the areas of large
      centrality and a steeper decay of density towards the periphery,
      as shown in fig.~\ref{oned}a. This can intuitively be understood
      by looking {\it e.g.} at the role played by the parameter
      $\beta$ in Eq.~(\ref{oned}): as $\beta$ decreases to 0, the
      differences in rate of growth in different locations becomes
      negligible }
  \item{Eq.~(\ref{eq:oned}) describes the average (or expected)
      behavior of the population density over time.  Numerical
      simulations of the corresponding stochastic process show
      fluctuations from the above mentioned expected value. Such
      fluctuations increases as noise increases (ie. when $\beta$ decreases).}
  \item{Numerical integration of Eq.~(\ref{eq:oned}) suggests that, as
      $\lambda$ increases, the decay of density assumes a power law
      form whose exponent depends on $\beta$ and $\lambda$ and
      approaches $-1$ as $\lambda$ gets very large. This can be
      explained by assuming $f(x) \approx \gamma x^{-r} $, and using
      Eq.~(\ref{onedeq}) and its derivative both computed in $L$. The
      derivative of $f(x)$ is:
\begin{equation}
f'(x)=\beta \lambda f(x)\left( 1- 2 \int_{-L}^{x}f(y)dy \right) f(x).
\end{equation}
The above expression can be computed in $x=L$, taking into account
that that $\int_{-L}^{L}f(x)dx=1$ and the assumed algebraic functional
form of $f(x)$. This leads to
\begin{equation}
\gamma r L^{-1} = \gamma \beta L^{-r}( \lambda \gamma - r \gamma L^{-1}). 
\end{equation}
In the limit of large $L$ one can keep only the leading orders in $L$,
and match the power and the coefficient of the leading order on the
two sides of the equation above.  This gives $r=1$ and $\gamma=
1/(\beta \lambda)$. The validity of this argument can be verified
looking at fig. \ref{oned}b, where $\ln(f)$ is plotted vs $\ln(x)$ to
highlight the power law behavior of $f(x)$ and where the line $1/
(\beta \lambda x)$ has been plotted as a reference for the case $\beta
= 10$ and $\lambda=10$.}

\end{itemize}

This simple one-dimensional model thus allowed us to understand some basic
features of the model that will be discussed in their full generality in
the next section.

\subsection{Two-dimensional case:  existence of a localized regime}

\begin{figure}
  \centering
    \includegraphics[angle=0,scale=.33]{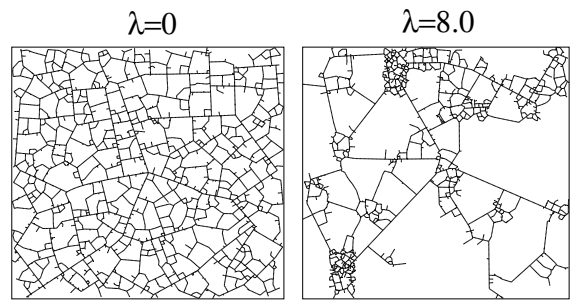}
    \caption{Networks obtained for different values of $\lambda$ (and
      for $N=500$ and $\beta=1$). On the left, $\lambda=0$ and
      only the density plays a role and we obtain a uniform
      distribution of centers. On the right, we show the network
      obtained for $\lambda=8$. In this case, the centrality is the
      most important factor leading to a few dominant areas with high
      density.}
   \label{fig:lambda}
\end{figure}

We now apply the probability in Eq.~(\ref{eq:proba}) to the growth
model described in the first part of this paper.  The process starts
with a `seed' population settlement (few centers distributed over a
small area) and a small network of roads that connects them. At any
stage, the density and the betweenness centrality of all different
subareas are computed, and a few new centers are introduced. Their
location in the existing subareas is determined according to the
probability defined in Eq.~(\ref{eq:proba}). Roads are then grown
until the centers that just entered the scene are connected to the
existing network. This process is iterated until the desired number of
centers has been introduced and connected.  In the two panels of
figure~\ref{fig:lambda} we show the emergent pattern of roads that is
obtained when $\lambda$ is small and very large, respectively.

When $\lambda$ is small the density plays the dominant role in
determining the location of new centers. New centers appear preferably
where density is small, smoothing out the eventual fluctuations in
density that may occur by chance and the resulting density is
uniform. On the other hand, when $\lambda$ is very large, centrality
plays the key role, leading to a city where all centers are located in
the same small area.  The centrality has thus an effect opposite to
that of density and tends to favor concentration. We will now describe
in more details the transition between the two regimes described
above.

We compute, in the two cases, the following quantity (previously
introduced in a different
context~\cite{Derrida:1987,Barthelemy:2003b}):
\begin{equation}
Y_2=\sum_i\left[\frac{N(i)}{N}\right]^2
\end{equation}
where the sum runs over all sectors which number is $N_s$. In the uniform case, all
the $N(i)$ are approximately equal and one obtains $Y_2\sim 1/N_s$,
which is usually small. In contrast, when most of the population
concentrates in just a few sectors which represent a finite fraction
of the total population, we obtain $Y_2\sim 1/n$ where $n$ represents
the order of magnitude of these highly-populated sectors- the
`dominating sectors'. The quantity
\begin{equation}
\sigma=\frac{1}{Y_2N_s}
\end{equation}
gives therefore the fraction of dominating sectors.
\begin{figure}
  \centering
    \includegraphics[angle=0,scale=.25]{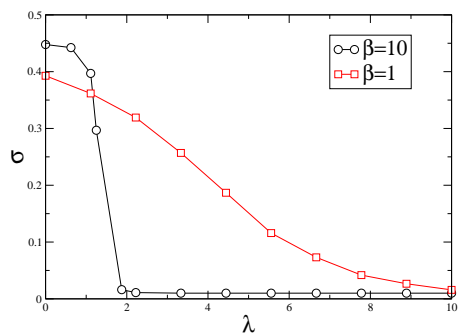}
    \caption{Fraction of dominating sectors (obtained for $500$
      centers and averaged over $100$ configurations). When $\lambda$
      is small, the center distribution is more uniform and $\sigma$ is
      large (close to $100\%$).When $\lambda$ increases, we see the
      appearance of a few sectors dominating and concentrating most of
      the population. This effect is smoothen out for smaller values
      of $\beta$ corresponding to the possibility of choice.  }
   \label{fig:fraction}
\end{figure} 
The behavior of $\sigma$ vs. $\lambda$ is shown in
Fig.~\ref{fig:fraction}.  We observe that $\sigma$ decreases very fast
when $\lambda$ increases, signaling that a phenomenon of
localization sets in as soon as transportation costs are involved.

We conclude this section discussing the role played by the parameter
$\beta$.  Analogously to what happens in the one dimensional case, the
concentration effect is weakened by a small values of $\beta$.  The
parameter $\beta$ describes the overall importance of the cost-factors
with respect to other factors that have not been explicitly taken into
account, or, equivalently, the possibility of choice.  Indeed, when
$\beta$ is very large, the location which maximizes the cost is
chosen. In contrast, when the parameter $\beta$ is small, the cost
differences are smoothen out and a broader range of choices is
available for new settlements. Figure~\ref{fig:fraction} illustrates
the importance of choice. In particular, the appearance of
large-density zones (controlled by the importance of transportation
accessibility) is counterbalanced by the possibility of choice and the
resulting pattern is more uniform.

\subsection{Density profile: the appearance of core districts}

\begin{figure}
\vspace*{1.0cm}
  \centering
    \includegraphics[angle=0,scale=.30]{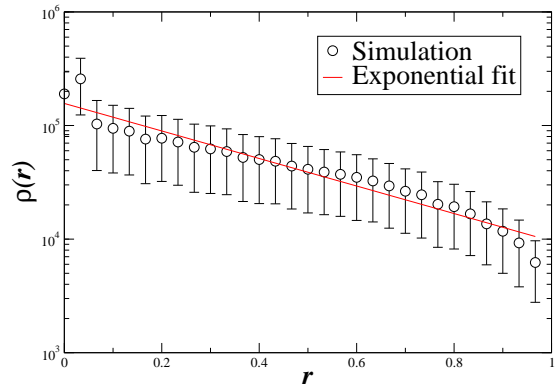}
    \caption{Density profiles for $\lambda=8$ ($N=200$, averaged over
      $500$ configurations). The decay of the density profile is well
      fitted by an exponential, signalling the appearance of a
      well-defined core district (error bars represent one standard
      deviation). }
   \label{fig:density}
\end{figure}

In this last part, we describe the effect of the interplay of
transportation and rent costs on the decay of population density from
the city center. In the following, the core district is identified
as the sector with the largest density. The whole plane is then
divided in concentric shells with internal radius $r$ and width
$dr$. The density profile $\rho (r)$ is given by the ratio of the
number $\delta n$ of centers in a shell to its surface $\delta S(r)$
\begin{equation}
\rho(r)=\frac{\delta n}{\delta S}
\end{equation}
For small $\lambda$, the density is uniform, as expected.
In figure \ref{fig:density} we show the density profile $\rho(r)$ in
the case of $\lambda$ large,
where we observe a fast exponential decay of the form $\exp{-r/r_c}$,
in agreement with empirical observations \cite{Makse1}.  This behavior
is the signature of the appearance of a well-defined core district of
typical size $r_c$, whose typical size $r_c$ decreases with $\lambda$.
This simplified model predicts, therefore, the existence of a highly
populated central area whose size can be estimated in terms of the
relative importance of transport and rent costs.

%%%%%%%%%%%%%%%%%%%%%%%%%%%%%%%%%%%%%%%%%%%%%%%%%%%%%%%%%%%%
\section{Discussion and perspectives}
% ---------------- Conclusion

We presented a basic model that describes the impact of economical
mechanisms on the evolution of the population density and the topology
of the road network. The interplay between rent costs and demand for
accessibility leads to a transition in the population spatial
density. When transportation costs are moderate, the density is
approximately uniform and the road network is a typical planar network
that does not show any strong heterogeneity. In contrast, if
transportation costs are higher, we observe the appearance of a very
densely populated area around which the density decays exponentially,
in agreement with previous empirical findings. The model also predicts that
the demand for accessibility easily prevails on the disincentive
constituted by high rent costs.

A very important ingredient in modeling the evolution of a city is how
individuals choose the location for a new business or a new home. We
isolated in this work the two important factors of rent price and
transportations costs. For these costs, we assume some reasonable
forms but it is clear that large scale empirical measures are
needed. In particular, it would be interesting to characterize
empirically how the rent price varies with the density and how
transportation costs varies with the centrality. Possible outcomes to
these studies would be to give an idea of the value of the parameter
$\lambda$ (and possibly also $\beta$) and thus to determine how much
the city is centralized. 

As it happens in every modeling effort, a satisfactory compromise
between realism and feasibility must be found and we opted for
sacrificing some important economic considerations in order to be able
to explicitly take into account the topology of the road
transportation network and not the distance to a center only, as it is
usually assumed in most models. Our model predicts so far the
appearance of a core center, but it is however known that cities
present a large diversity in their structure, ranging from a
monocentric organization to different levels of polycentrism. In
addition, interesting scaling relations between different parameters
(total wages, walking speed, total traveled length, etc) and
population size were recently found
~\cite{Bettencourt:2007,Moses:2007} showing that beyond the apparent
diversity, there are some fundamental processes driving the evolution
of a city.

These different results appear as various facets of the process of
city formation and evolution and it is at this stage not clear how to connect the
scaling to the structural organization of a city and more generally,
how to reconciliate the different existing results in a unified
picture. We believe that the present model- modified or generalized-
could help for future studies in this direction.

%%%%%%%%%%%%%%%%%%%%%%%%%%
Acknowledgments. We thank G. Santoboni for many discussions at various
stages of this work, and two anonymous referees for several important
comments and suggestions. MB also thanks Indiana University for its
warm welcome where part of this work was performed.

%%%%%%%%%%%%%%%references

%%%%%%%%%%%%%%%%%%%%%%%%%%%%%%%%%%%%%%%%%%%%%%%%%%%% 

\end{document}